# REGULARIZATION OF TUNNELING RATES WITH QUANTUM CHAOS

LOUIS M. PECORA
*Materials Physics and Sensors*
*U.S. Naval Research Laboratory*
*Washington, DC 20375, USA*
*pecora@anvil.nrl.navy.mil*

HOSHIK LEE
*Department of Physics*
*College of William and Mary*
*Williamsburg, VA, USA*
*hoshik.lee@wm.edu*

DONG-HO WU
*Materials Physics and Sensors*
*U.S. Naval Research Laboratory*
*Washington, DC 20375, USA*
*dhwu@ccs.nrl.navy.mil*

We study tunneling in various shaped, closed, two-dimensional, flat potential, double wells by calculating the energy splitting between symmetric and anti-symmetric state pairs. For shapes that have regular or nearly regular classical behavior (e.g. rectangular or circular) the tunneling rates vary greatly over wide ranges often by several orders of magnitude. However, for well shapes that admit more classically chaotic behavior (e.g. the stadium, the Sinai billiard) the range of tunneling rates narrows, often by orders of magnitude. This dramatic narrowing appears to come from destabilization of periodic orbits in the regular wells that produce the largest and smallest tunneling rates and causes the splitting vs. energy relation to take on a possibly universal shape. It is in this sense that we say the quantum chaos regularizes the tunneling rates.

## 1. Introduction

A general theory of tunneling between two potential wells separated by a barrier has been well known for many decades for one-dimensional systems [Cohen-Tannoudji et al., 2006]. See Fig. 1. However, in two-dimensional systems a general theory of tunneling "has largely remained impervious to such analysis." [Creagh & Whelan, 2001]. A similar statement also applies to opens systems where metastable states govern behavior [Creagh & Whelan, 1998]. This makes the discovery of universal behaviors important to guide theory and, in the absence of a general theory, guide intuition about double well potentials, especially for their implementation in devices like quantum dots and sensors [Loss and Divincenzo, 2006].

Some recent work has tied tunneling rates in chaotic or mixed systems to (perhaps, unstable) periodic orbits in a semiclassical realm [Creagh & Whelan, 2001] and [Creagh & Whelan, 1996; Smith & Creagh, 2006; Tomsovic & Ullmo, 1994] and in relation to wavefunction scarring [Bies et al., 2001]. In those studies the barrier was often kept within a small region of the potential or particular periodic orbits were used. While these led to some insights for systems with such barriers, they could say little about long barriers or more complicated orbital situations. This is not a criticism, but an acknowledgment of the difficulty of the problem.

We examine several two-dimensional, flat-potential, double-well systems in which the barriers run along most of one side of each well. We numerically solve Schrödinger's equation (in the Helmholtz form) for several hundred eigenstates in each system. The differences in energies between the symmetric and antisymmetric pairs of states give the tunneling rates for wavefunctions in combinations of those states. For integrable or near integrable wells the tunneling rates can vary by several orders of magnitude for states that are very close in energy. This is unlike the one-dimensional case (Fig. 1) where tunneling rates monotonically increase with energy. However, as we move to systems with ubiquitous chaos, the variation of the tunneling rates with energy decreases to a narrow range and the graphs of tunneling rates vs. energy approach what appears to be a universal curve resembling in some ways the curve in the one-dimensional situation. It appears that the destruction of entire sets of stable or marginally stable periodic orbits by changes in boundary shapes causes this change in tunneling rates. Using Husimi distributions to get the average probability and average momentum normal to the barrier ($p_x$ in most of our systems) shows that the tunneling rates are closely related to the size of the wave function at the barrier and the magnitude of the momentum normal to the barrier.

## 2. Eigenstates and Tunneling Rates

We have chosen six well shapes: rectangular, circular, stadium, Sinai, butterfly, and concave. We use each shape to form two wells separated by a long barrier which is straight on each side (or nearly so), except for the circular wells. Their geometry can be seen in Fig. 1. The areas of all wells are equal (=4.8 in arbitrary units) so that in an average sense (using Weyl's formula [Gutzwiller, 1990] as a guide) the eigenstates will be similarly distributed in energy. The barriers are all of the same average width (0.1 arb.units) and potential height ($V_b$ =1000) so the splittings will roughly cover the same range. In this way we can fairly compare the tunneling rates of different shapes.

To find the stationary eigenstates of the double well systems we employ the boundary element method (BEM) [Knipp & Reinecke, 1996; Ram-Mohan, 2002] to solve Schrödinger's equation in the Helmholtz form, $\left(\nabla^2 + k^2\right)\psi(\mathbf{r}) = 0$, where $k = \sqrt{E-V}$, $E$ is the eigenstate energy and $V$ is the potential level of the well (taken as $V$=0 here). We take $\hbar^2/2m = 1$ so all energies are in inverse-length-squared units. We note that we can find energy level differences to within less than 1% accuracy. For the Sinai system we used the finite element method.

We calculated all eigenenergies and states from the ground state up to an energy of approximately 800 (just below the barrier potential). This gives about 300 states per double-well system or about 150 symmetric-antisymmetric pairs. Fig. 1 shows the tunneling rates for each system as a function of the mean energy of the splitting pairs along with some typical eigenstates.

In Fig. 1 (c) the tunneling rates depend directly on the momentum normal to the barrier since $p_x$ and $p_y$ are good eigenvalues labeling each state and only the $p_x$ value affects the tunneling rate. This results in "lines" of equal valued tunneling rates for states that all have the same $p_x$ value, but different $p_y$ values and, hence, energies. The tunneling rates for eigenstates of almost equal energy can be vastly different by ratios of over two orders of magnitude.

Fig. 1 (d) shows a similar range of tunneling rates with non-monotonic energy dependence. Even though $p_x$ and $p_y$ are not good eigenvalue labels for the states, the system is nearly integrable and the eigenstates resemble those of the simple circular well (Bessel functions times sinusoids). The eigenstates are highly structured giving some patterns with very low or very high probability near the barrier and have high variation in momentum normal to the boundary. This combination leads again to large variations in tunneling rates for states close in energy.

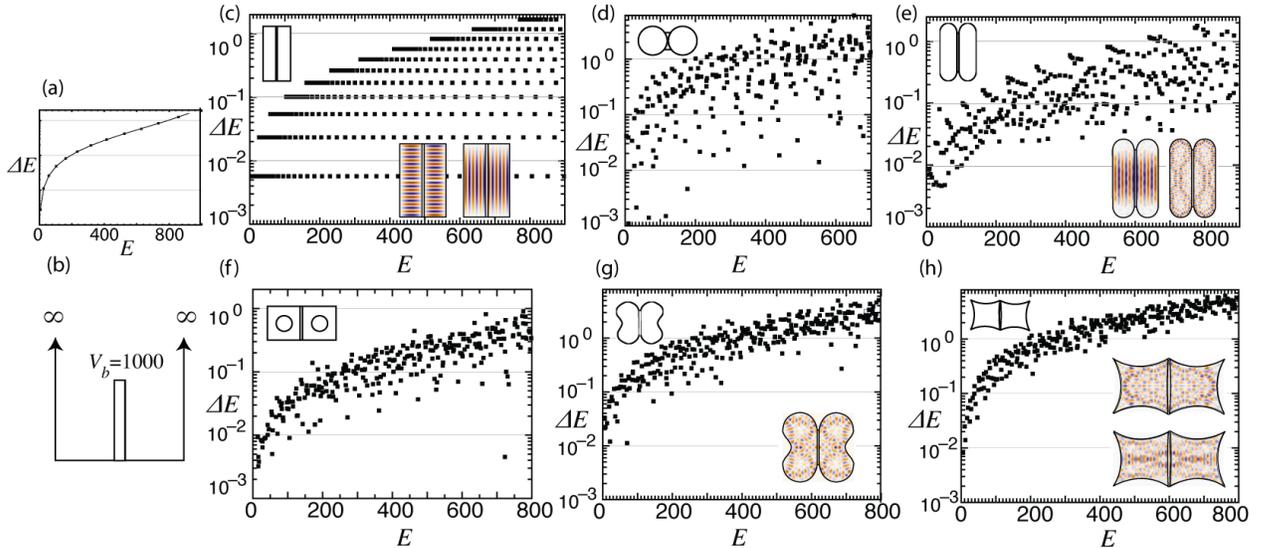

Fig. 1 (a) and (b) One dimensional wells and the associated tunneling rates vs. energy. Tunneling rates for (c) rectangular, (d) circular, (e) stadium, (f) Sinai, (e) butterfly, and (h) concave double-well systems. Insets show the double-well and barrier configuration (not to scale) and some typical eigenstates. For the rectangular well (c) low tunneling and a high tunneling rate states are shown.

In Fig. 1 (e) for a double stadium system we begin to see changes in the variations of tunneling rates. Notably absent in the higher energy states (>150) are the smallest rates common in both (c) and (d). Some remanence of the lines of higher tunneling rates for the rectangular wells are still evident. Fig. 1 (f) for the Sinai double wells shows more contraction of the energy variation beyond that of the stadium case. Gone are the high lines of tunneling rates. For the butterfly shaped system of Fig. 1 (g) we see further diminishing of tunneling rate variations for nearby energies. And, finally, for the case of fully concave potential wells in Fig. 1 (h) we have the minimum variation of tunneling rates – approximately within a factor of 2 of neighboring states – a decrease of two orders of magnitude from Fig. 1 (c) and (d).

## 3. Poincaré Sections

We can get some understanding of the tunneling rates by examining the classical dynamics of single billiards. In Fig. 2 we plot bounce or bounce maps or Poincaré sections in Birkhoff coordinates [Nakamura & Harayama, 2004] of single wells using the variables $s$ the distance along the perimeter of a well to where a collision with the wall occurs for a billiard particle and $c$ the sine of the angle of incidence at the collision which capture the dynamics of a single billiard well [Nakamura & Harayama, 2004]. The maps were made by selecting 400 random initial conditions in the phase space $(s,c)$ and iterating for 400 collisions each. This is a sampling of the dynamics of the well, not a rigorous search for various orbits or regions.

In Fig. 2 (a) the Birkhoff map is made up of horizontal line segments. This is one sign of an integrable (or nearly so) well. The lines are generated by bounces back and forth between opposite walls. In the quantum realm this leads to states well-labeled by $p_x$ and $p_y$. Fig. 2 (b) shows similar horizontal trajectory sets which come from orbits that bounce around the circle from straight-across bounces to whispering gallery modes. These lead in the quantum case to very nearly integrable states which like in the rectangular system have extreme values of momenta normal to the boundary. (Strictly speaking the circular double well is probably not integrable, but is very close to an integrable system since the barrier is rather high and even high energy states closely resemble the isolated circular well states.) Fig. 2 (c) for the stadium well is best compared with the rectangular well. We note that most of the horizontal lines are gone. This is an ergodic, strongly chaotic system [E. Ott, 1995], but we notice that the bounce orbits between left and right walls are not very unstable leading to a lot of horizontal groups of points along $c=0$. This leads to states where $p_x$ is almost a good state label which is similar to the rectangular well (see Fig. 1). However, the top-bottom family of bounce orbits (which results in the horizontal-striped quantum state) are not present and the associated states in rectangular (see Fig. 1) never show up in the stadium. They have been destabilized by the rounded ends of the stadium. Hence, this is why we see the rows of high tunneling rates for the rectangular and stadium systems, but not similar low tunneling rates for the stadium. Fig. 2 (d) shows that the central circle in the Sinai billiard mostly destabilizes the side-to-side and top-to-bottom orbits although some still remain which avoid the circle and their signature can be seen as small horizontal point sets along the center of the map. This change results in destabilization of the strong $p_x$ states still present in the stadium system and kills off the rows of high tunneling rates. Such orbits probably prevent the tunneling rates from narrowing more in variation, but we have to investigate this in more detail. In Fig. 2 (e) we see that the butterfly well with its indented side has completely destabilized the side-to-side orbits leading to a map with almost no structure and smaller tunneling rate variations. This trend is continued in Fig. 2 (f) where no structure is apparent and the tunneling rate variation is narrowest of all. The continually curved boundaries in this last case sufficiently destabilize all orbits leading to the narrowest of all tunneling rate curves.

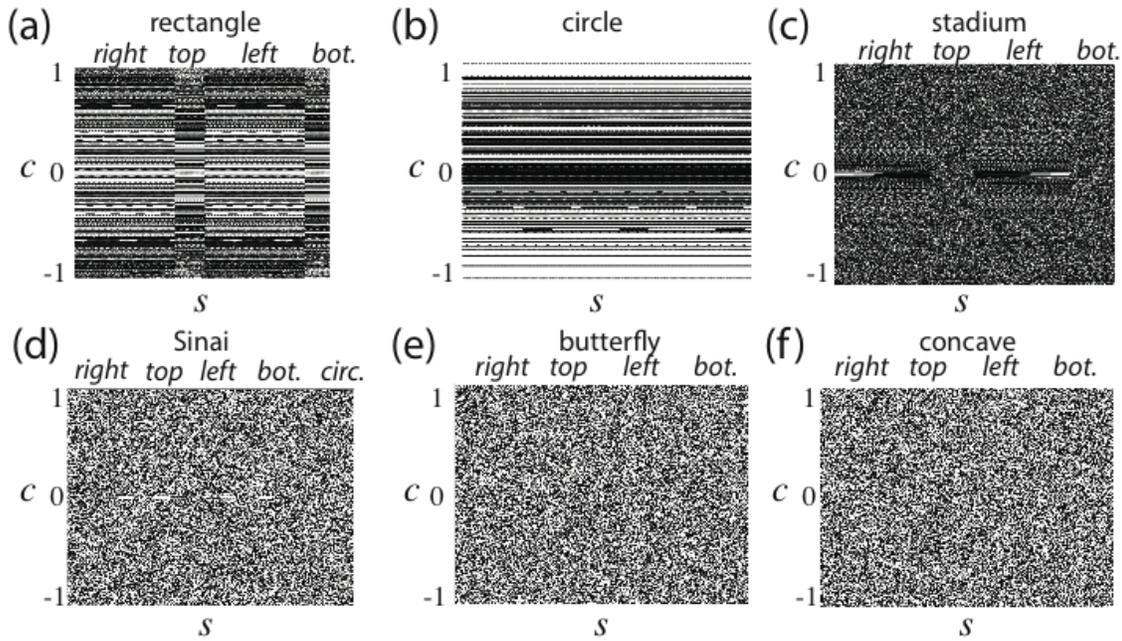

Fig. 2 Poincaré sections for the six wells (a) rectangular, (b) circular, (c) stadium, (d) Sinai, (e) butterfly, and (f) concave.

## 4. Tunneling Rates and Momentum at the Barrier

In order to get a more quantitative analysis we projected the wave function from each state onto a coherent state [Crespi et al., 1993; Luna-Acosta et al., 1996] at the barrier. This Husimi distribution provided us with an estimate of the distribution of the wave function along the barrier ($y_0$) and the momentum parallel to the barrier ($p_y$), $H(y_0, p_y)$ [Crespi et al., 1993; Luna-Acosta et al., 1996]. From this we calculated the average magnitude (probability) at the barrier $\langle \psi \rangle \equiv \iint H(y_0, p_y) dy_0 dp_y$ and the average momentum normal to the barrier $\langle p_x \rangle \equiv \iint \sqrt{E - p_y^2} H(y_0, p_y) dy_0 dp_y / \langle \psi \rangle$. Individual plots of tunneling rates vs. these quantities show the best correlation comes from plotting tunneling rates vs. weighted average normal momentum = $\langle \psi \rangle \langle p_x \rangle$ (see Fig. 3). This makes physical sense and gives at least a semi-quantitative characterization of the tunneling properties of the well systems we studied.

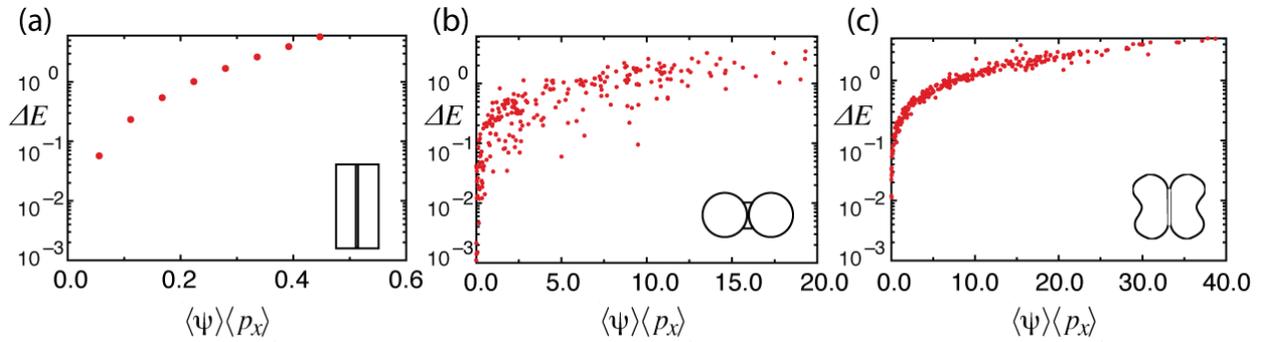

Fig. 3 Tunneling rates (energy splittings) for (a) rectangular, (b) circular, and (c) butterfly double-well systems vs. average weighted normal momentum to the barrier $\langle \psi \rangle \langle p_x \rangle$. Other chaotic well systems are very similar to (c).

Fig. 3 confirms the qualitative view we had about tunneling rates: tunneling rates depend on the probability of the particle being at the barrier and its momentum normal to the barrier. Note that in Fig. 3 and Fig. 1 the tunneling rates are in roughly one-to-one correspondence with both the energy and the weighted normal momentum to the barrier for the chaotic systems implying that the energy strongly predicts the weighted momentum, too. This is not true for the integrable systems.

## 5. Conclusions

At present a simple theory that explains this phenomena of tunneling rate regularization is absent from the literature [Creagh, 2009; Fishman, 2009]. It may be possible to use complex paths or instanton approaches to get semiclassical approximations to the tunneling in these systems, but these involve the more difficult situation where one must deal with dense families of periodic orbits rather than a few isolated orbits [Creagh, 2009; Fishman, 2009]. Both of these ideas will take some effort at development and we hope to examine those possibilities in the future. We have found that a more productive approach is to approximate the form of the wave function in the wells with more chaotic and ergodic classical dynamics as a random plane wave superposition. This is under development and will be presented elsewhere [Pecora, et al., 2011].

A possible experimental consequence for similar real systems may be that since injection of electrons into nanowells is rarely monochromatic, tunneling currents may appear noisier in wells with regular shapes where even a narrow energy window encompasses vastly different tunneling rates than in chaotic wells.

On a more practical level the results here show that it may be possible to engineer well shapes that constrain and/or guarantee certain tunneling rates in some windows of the energy spectrum. One could now change tunneling rates using well shape rather than just by barrier height adjustment. The latter would not cure the problem of having high tunneling rate variability whereas the well shape would. That surely would be useful for any devices which operate just below or in the semiclassical limit. Certain 2DEG quantum dot and graphene systems already operate in this regime.

We have presented strong evidence that for double well systems which are classically strongly chaotic tunneling rates for quantum states are regularized to have very little local fluctuations with energy. This contrasts with integrable systems which appear to suffer from large fluctuations in tunneling rates. It will be interesting to see if this phenomenon carries over in any way to mixed systems and more general potentials.

We would like to thank the following people for helpful conversations and advice: P. Knipp, L. Ramdas Ram-Mohan, T. Reinecke, S. Creagh, M. White, and S. Fishman.